\documentclass[pdflatex,sn-mathphys-ay]{sn-jnl}

\usepackage{graphicx}%
\usepackage{epstopdf}
\usepackage{multirow}%
\usepackage{amsmath,amssymb,amsfonts}%
\usepackage{amsthm}%
\usepackage{mathrsfs}%
\usepackage[title]{appendix}%
\usepackage{xcolor}%
\usepackage{textcomp}%
\usepackage{manyfoot}%
\usepackage{booktabs}%
\usepackage{algorithm}%
\usepackage{algorithmicx}%
\usepackage{algpseudocode}%
\usepackage{listings}%
\usepackage{anyfontsize}%

\theoremstyle{thmstyleone}%
\theoremstyle{thmstyletwo}%

\theoremstyle{thmstylethree}%

\begin{document}

\title[Ethics of Blockchain Technologies]{Ethics of Blockchain Technologies}


\author[1]{\fnm{Georgy} \sur{Ishmaev}}\email{georgy.ishmaev@inria.fr}



\affil[1]{\orgdiv{Inria/Irisa}, \orgname{University of Rennes}, \orgaddress{\city{Rennes}, \country{France}}}




\abstract{This chapter explores three key questions in blockchain ethics. First, it situates blockchain ethics within the broader field of technology ethics, outlining its goals and guiding principles. Second, it examines the unique ethical challenges of blockchain applications, including permissionless systems, incentive mechanisms, and privacy concerns. Key obstacles, such as conceptual modeling and information asymmetries, are identified as critical issues. Finally, the chapter argues that blockchain ethics should be approached as an engineering discipline, emphasizing the analysis and design of trade-offs in complex systems.\footnote{Pre-print. \textit{Handbook of Blockchain and Society. L.Robb \& J.Flood (Eds.). Forthcoming.}}}

\maketitle

\section{Introduction}\label{sec1}

These days, discussions about the ethics of emerging technologies are common, with particular focus on areas like AI, private data collection, and autonomous vehicles. It seems straightforward to claim that as new technologies raise moral issues, it makes sense to explore their ethics. Similarly, blockchain technology, which can also present moral dilemmas and embody non-technical values, warrants an examination of its ethical dimensions.

However, determining what constitutes blockchain ethics is more complex. First, we must ask what outcomes we expect from blockchain ethics, what is its purpose? Second, defining the proper scope of this ethical inquiry is challenging, as blockchain is not a singular technology but an open-ended stack with diverse applications. Finally, we must consider how to practice blockchain ethics effectively; if ethical considerations do not impact the subject matter, their value is questionable. This chapter seeks to address these questions and discuss the associated challenges.

To begin, it is useful to examine the current state of blockchain research and industry to understand why we need blockchain ethics at all. Unlike with the aforementioned sub-fields of ethics of emerging technologies, it is fair to say that blockchain ethics is far from being a mature discipline. A review of the literature reveals that academic work on blockchain ethics is sparse, often limited to general observations~\cite{agerskov_ethical_2023}. This contrasts sharply with AI ethics, which has a wealth of specialized studies focusing on specific issues and use cases~\cite{laine_ethics-based_2024}.

The state of arts in the blockchain industry is even more instructive. Unlike AI ethics, which has permeated the industry and even become a field of its own, blockchain lacks systematic ethical discussions. \footnote{This is not a claim that because of the abundance of ethics discourse AI is a more ethical industry. In fact, the claim can be made that there is little correlation~\cite{munn_uselessness_2023}} The industry’s moral reputation is tarnished by events like the collapse of the FTX cryptocurrency exchange~\cite{conlon_collapse_2023}, frequent hacks, and a general perception of blockchain as a "decentralized casino".\footnote{\url{https://www.theblock.co/post/290892/the-memecoin-casino-undermines-cryptos-long-term-prospects-a16z}} While it might be unfair to judge the ethics of blockchain technology based solely on business practices, it is clear that there are significant tensions between various norms, values, and ethical principles in this space.\footnote{\url{ https://polynya.mirror.xyz/ptscXuh3J3KOj2uJAn0vrEanpn2nauwA7iytYZ4cM9U}}.

This situation is particularly discouraging given the morally driven origins of decentralized applications. Early experiments in digital currencies, like Bitcoin, were heavily influenced by the cypherpunk movement, which combined libertarian ideals with techno-optimism, framing the development of cryptographic tools as a moral endeavor~\cite{swartz2018bitcoin}. However, many current blockchain applications seem to have strayed from these original aspirations. Timothy May, author of the "Crypto Anarchist Manifesto", already back in 2018, expressed skepticism about the blockchain industry’s adherence to its philosophical roots.\footnote{\url{https://www.coindesk.com/markets/2018/10/19/enough-with-the-ico-me-so-horny-get-rich-quick-lambo-crypto/}}

This apparent contradiction makes blockchain ethics an intriguing field of study, but it also raise a question whether actionable ethical approaches are even feasible at all. Some may adopt a radically skeptical stance, suggesting that blockchain ethics is a futile undertaking. However, this perspective is not helpful for several reasons. To claim that blockchain solutions can never be morally desirable is a form of naive defeatism, similar to "AI doomerism," which argues that the only morally acceptable solution is to cease developing applications entirely.

Firstly, such a stance is unrealistic. Even in a fantastic scenario where governments around the world were to reach a consensus to ban blockchain technologies, history shows that no technology has ever been successfully eradicated. More likely, such technologies would be co-opted by powerful actors who would then define the values for their users, as seen in how Facebook has redefined "privacy"~\cite{boyd_facebook_2010}. And the failed launch of the "Libra" could already be seen as such an attempt~\cite{abraham2019other}.The second argument against defeatism is that abandoning blockchain ethics means forfeiting potential benefits of these technologies. Furthermore, decentralized technologies are rarely value-neutral, blockchains, in particular, are not neutral because they are also systems based on incentives mechanisms that shape human behavior in particular ways. Thus, we can not simply sweep away the questions of normative character.

So, rather than succumbing to defeatism, there are good reasons to try and develop systematic approaches to the ethics of blockchain applications. This chapter will address the three key questions in blockchain ethics. Section \ref{sec:eth_tech} will explore the place of blockchain ethics within the broader context of technology ethics and define its goals and purposes. Section \ref{sec:block_eth} will examine what makes blockchain and decentralized technologies unique, helping to define the appropriate scope of blockchain ethics. Section \ref{sec:obstacles} will identify the challenges and obstacles facing blockchain ethics as both a research field and a practical endeavor. Finally, Section \ref{sec:pract_eth} will argue that the most promising way to make blockchain ethics feasible is to approach it as an engineering discipline that helps us understand and design trade-offs in complex systems.

\section{Ethics of technology}
\label{sec:eth_tech}

When discussing blockchain ethics, it is crucial to clarify what do we mean by "ethics." While the concept may seem straightforward, it is important to gain a precise understanding in order to appreciate it in different contexts. Fortunately, we can draw from the extensive history of moral philosophy, a branch of philosophy with a venerable tradition nearly as old as philosophy itself. In this tradition, ethics is commonly defined as the systematic study of moral matters. This may seem a bit circular but what it is means is that ethics tries to address such questions as right and wrong, moral norms, values and how should we treat each other. What also distinguishes ethics is its normative nature, it doesn’t merely describe moral beliefs but evaluates them to help make moral decisions.

Moral philosophy encompasses various disciplines, but for our purposes, it is useful to distinguish between three broad fields: \textit{metaethics, normative ethics}, and \textit{applied ethics}. These fields can be arranged from higher to lower levels of abstraction. \textit{Metaethics} addresses foundational issues, such as whether there are moral truths and how we can know them (or not) (For good introduction see~\cite{chrisman_what_2023}). \textit{Normative ethics} deals with constructing and explicating moral principles, including well-known theories like deontology, consequentialism, and virtue ethics ~\cite{timmons_moral_2012}. \textit{Applied ethics}, in turn, evaluates whether specific actions in particular contexts are morally desirable. Examples include bioethics, business ethics, and technology ethics, the latter of which addresses moral issues in technological contexts, focusing on the development, implementation, and use of specific technologies~\cite{van_de_poel_ethics_2023}.

Understanding these distinctions is essential to avoid unwarranted generalizations and to identify the appropriate level of abstraction for the problems we are addressing. For instance, practitioners might be disappointed that there is no single, coherent theory of ethics that can immediately resolve specific moral dilemmas. More abstract theories often underdetermine\footnote{E.g. a moral theorist working on specific branch of consequentialism may sometimes assume an affirmative answer to some metaethical problems.} lower-level investigations, and normative theories do not easily translate into actionable moral guidelines in specific contexts.

This challenge is particularly evident in ethics of technology, where normative theories often have limited explanatory value in identifying and resolving immediate moral intuitions~\cite{hansson_moral_2023}. This may be indeed disappointing, but on some rational reflection it becomes clear that it is unfair to expect that Kant would have envisioned and accommodated extension of his theory to address the resolution of moral conflicts in memecoins. Instead, we often have to do the "heavy lifting" ourselves in specific technological contexts. This involves identifying salient technological facts, translating them into the appropriate conceptual level of moral reasoning, and addressing distinctively moral issues that cannot be resolved by descriptive investigations alone.

Thus, technology ethics is not a simple top-down application of ready theories to specific cases~\cite{van_den_hoven_moral_2008}. A top-down approach often leads to oversimplification and confirmation bias, akin to the "spherical cow in a vacuum" analogy in science, where arguments are based on overly simplified models rather than actual systems\footnote{\url{https://en.wikipedia.org/wiki/Spherical_cow_in_a_vacuum}}. This issue frequently arises in humanities studies of blockchain solutions, where discussions too often revolve around inaccurate general conceptions of these systems rather than discussing actual systems.

For example, consider a hypothetical blockchain scaling solution that aims to improve performance. We might ask about the measurable performance increase, such as transactions per second, and the security guarantees of this solution - questions that can be empirically addressed. However, if we inquire whether these changes benefit only a limited set of blockchain node operators or a wider set of network users, we enter philosophical territory that cannot be readily addressed with experiments. Questions of fairness, for instance, are ethical issues that require more than purely empirical measurements. Addressing these questions requires meaningful engagement with normative theories of fairness, and adaptation of these insights to specific technological contexts.

Once empirical observations are exhausted, we can analyze the problem using relevant theories. Normative theories offer conceptual tools to express moral issues more precisely than everyday language and provide valid reasoning patterns for moral norms and judgments~\cite{van_den_hoven_moral_2008}~\cite{hansson_theories_2017}). Continuing with our example, if we determine that the salient ethical concern is fairness, we should analyze our context-specific insights within an appropriate ethical framework on fairness.

In summary, blockchain ethics should, at a minimum, help us identify and explain distinctively ethical issues that cannot be resolved through empirical observations alone. Once these issues - whether value conflicts, moral dilemmas, risks, harms, or other concerns - are identified, they should be analyzed within the appropriate conceptual framework. However, identifying and explaining problems is not enough; we must address them meaningfully. Like ethics in general, blockchain ethics is not limited to identifying morally problematic issues but also seeks to understand how technology can promote human well-being and flourishing. But before addressing these questions, we need first to clarify the scope of "blockchain ethics" and determine what makes it distinctively different from other subfields of technology ethics.

\section{What is at stake in blockchain ethics}
\label{sec:block_eth}

To argue for the specific field of "blockchain ethics" is to argue that ethical questions presented by this research and practice are different enough from the already existing fields. Of course, moral philosophy itself is largely an open enterprise where anyone is free to propose "ethics of something". So this is not a strict requirement but it makes sense methodologically to identify whether we are dealing with qualitatively new concerns and problems that can not be satisfactorily explained in the existing conceptual frameworks of say more general ethics of information technologies. This section is not aiming to provide an exhaustive list of all areas of interest for blockchain ethics, as this list would be far beyond the scope of a single book chapter. Rather, we are going to look into three broader areas of ethical concerns that make blockchain ethics distinctive.

\subsection{Permissionless systems and identity}

To understand the concept of a "permissionless system," it is helpful to consider first how most modern computer systems are built. Most of these systems are distributed, meaning they consist of various devices that communicate and coordinate to achieve specific goals.\footnote{There is sometimes a confusion between terms "distributed" and "decentralize", as for example in the industry the former is used in the meaning of "even more decentralized". Here, for the sake of clarity we are using a definition of distributed systems that is conventional in computer science.} For example, a streaming service might appear as a single entity but is actually a network of data centers. The same applies to online shops, airline booking systems, banks, and pretty much any other modern computer system. All these systems face common challenges: determining which nodes should provide resources, which nodes can access them, and under what conditions~\cite{blaze_role_1999}.

The simplest way to address these challenges is by assigning identities to network participants and implementing access control based on these identities. This approach requires a trusted authority to assign identities and ensure that resources are accessible only to the authorized entities, creating a permissioned system where participants need permission to join, contribute, and consume resources.\footnote{"Permissioned" is usually used to highlight centralised blockchain solutions but here it is used in a general sense.} This principle is nearly universal, whether we consider computers in a data center or users of a streaming service, or in fact almost any other modern computer systems.

This, of course, makes a lot of sense. Random participants accessing random services or sharing random messages can easily cause chaos, not to mention potential for the malicious abuse of systems. Thus, most systems today rely on some form of identity-based access control, making coordination, security, and even time management in large distributed systems more manageable.

So, one may ask if there are good reasons to build and use such systems why do we even need alternatives? And why is this even an ethical question? And the answer is: as we get more dependent on those systems we get more dependent on trusted parties in these architectures. And sometimes this is not a good thing. Such trusted parties get to decide: who can join, who can participate, and even who is who. This may sound somewhat abstract, but in fact these are the problems behind some of the most discussed issues of recent years. For instance, platforms like Facebook decide what news we see, PayPal determines who can send money to whom, and GitHub controls who can share software. The list is extensive, and the issues are growing broader and more severe~\cite{van_dijck_platform_2018}.

The trajectory of computer systems evolution over the past decades has only amplified these concerns. In the 1980s, computer scientist David Chaum articulated key concerns about the impact of digital communication technologies, such as loss of privacy, autonomy, and the disempowerment of individuals as data increasingly concentrates in the hands of centralized entities~\cite{chaum_security_1985}. Chaum predicted that the logic of digitalization would lead to increased information asymmetries and power imbalances. More so, as Chaum has correctly explained, technological architectures may become blueprints shaping the social order.

A modern example is the use of COVID-19 credentials by the Chinese government to restrict protestors' movement, highlighting how centralized technical architectures can reduce humans to mere endpoints in a system. \footnote{\url{https://www.reuters.com/world/china/china-bank-protest-stopped-by-health-codes-turning-red-depositors-say-2022-06-14/}} This underscores the need for exploring alternative architectures that incorporate ethical concerns rather than subjugating human participants to the system's logic. Decentralized, permissionless solutions, that do not necessarily require strong persistent identifiers for participants offer an interesting alternative.

It is important, of course, not to fall into the trap of a technological determinism, as these issues also have institutional, economic, and political dimensions. However, technological solutions have inherent constraints and affordances that define their possible uses. Just as asymmetric encryption enables certain privacy and trust relations that symmetric encryption does not~\cite{chaum_security_1985}, permissionless systems enable relations that permissioned systems do not.

The original Bitcoin protocol elegantly addresses fundamental issues of distributed systems without requiring persistent identities or permissions. Instead of tracking miner identities to prevent misbehavior, Bitcoin only considers the amount of work performed by miners and rewards them accordingly. Similarly, users need only a valid cryptographic signature to transfer value, bypassing the need for persistent identities and trusted access-control authorities~\cite{cachin_blockchain_2017}.

This approach has significant ethical implications not just for computer scientists but for societies dependent on digital infrastructures. It demonstrates that systems can function on a significant scale without requiring participants' identification or centralized access control in a morally significant use cases. Not unlike the way in which encrypted communication tools are morally significant for the protection of privacy. For instance, cryptocurrencies have helped refugees preserve property while fleeing conflicts and have enabled people in inflation-prone regions to hedge their savings using blockchain-based stable-coins.\footnote{\url{https://www.cnbc.com/2022/03/23/ukrainian-flees-to-poland-with-2000-in-bitcoin-on-usb-drive.html}\footnote{\url{https://www.coindesk.com/policy/2022/09/13/latin-americans-turning-to-dollar-stablecoins-amid-inflation-surge-paxos/}}}

These examples show blockchain solutions functioning as originally intended: as "anti-dystopian" technologies. Thus, permissionless blockchains and similar decentralized systems can be seen as enabling technologies for human well-being. That is, they can enable (and preserve) moral values or norms that we find morally desirable in a way that other technological solutions can not.

\subsection{Privacy and Transparency}

Decentralized permissionless systems, as discussed above, offer a unique approach to designing distributed systems that reduce reliance on trusted parties and persistent identifiers, which has significant implications for privacy. Intuitively, minimizing reliance on persistent identifiers should enhance privacy. Moreover, in any system, technical privacy guarantees are limited by the need to rely on a goodwill of trusted parties.

To illustrate, consider a communication platform using end-to-end encryption (e2e). If the platform provider cannot access encryption keys, privacy relies solely on the strength of the e2e protocol—this is a trust-minimized solution. However, if the provider holds the keys, privacy is contingent on the provider's goodwill, rather than on encryption protocol~\cite{unger2015sok}. More generally, this is a principle of reducing reliance on a goodwill of trusted parties through provable privacy guarantee. This principle was evident in the different privacy approaches of contact tracing apps deployed during the COVID-19 pandemic, with trust-minmized solutions indeed providing better privacy guarantees~\cite{troncoso_deploying_2022}.

Accordingly, decentralized protocols in principle can serve as enabling technology for privacy-preserving solutions. Unfortunately, at the moment, most of the blockchain applications that we have are not privacy-preserving but rather make adversarial surveillance of its users even easier~\cite{zhang2019sok}. The original application of blockchain - Bitcoin - provides only weak pseudonymity guarantees and no confidentiality guarantees to its users. 

That is, their identity could be somewhat obfuscated thanks to the use of different wallet addresses that in principle are not directly correlated with any personally identifiable information. However, transactions between different addresses are almost always fully publicly available on a Bitcoin ledger unless they are obfuscated with some additional privacy-preserving techniques (e.g. transaction mixing). The same is true for almost all other largest (in terms of user adoption) blockchain protocols including Ethereum. Thus, once deanonymized, users inadvertently reveal all their past transactions and interactions.

So from a certain point of view, it is safe to say that for users whose real identity is connected with their wallet addresses, privacy guarantees provided by blockchain protocols are even worse than those of traditional financial systems. Thus exposing its users who can be easily targeted by cybercriminals or authoritarian governments. Furthermore, this is morally problematic from a more general perspective, given that financial privacy is quickly evaporating in traditional financial systems\footnote{\url{https://www.cbsnews.com/news/mastercard-credit-card-customer-data-sold/}}. Thus instead of providing a counterbalance to the trend of expanding financial surveillance blockchain applications risk only accelerating it.

Despite this, blockchain systems can still enable privacy-preserving applications through trust-minimization and cryptographic guarantees. Privacy-preserving cryptocurrencies like Monero use advanced cryptographic techniques to anonymize transactions~\cite{deuber2022sok}. Other projects aim to offer privacy-preserving smart contracts~\cite{qi2024sok}. However, technical challenges are not the only obstacles; regulatory environments often oppose privacy-preserving technologies, even in democratic regions, as seen in the prosecution of Tornado Cash developers.\footnote{\url{https://www.wired.com/story/tornado-cash-developer-found-guilty-of-laundering-crypto/}}

This regulatory challenge echoes the broader "crypto-wars" that began before blockchain, where governments reacted against asymmetric encryption tools. With each new generation of privacy-enhancing tools, this conflict resurfaces, as demonstrated by recent efforts to undermine encrypted messaging privacy. These issues are tied to broader political and philosophical debates, often driven by the vested interests of organizations and government offices in expanding surveillance, which conflicts with fundamental human rights. We will touch upon some of the regulatory challenges in more details in the Section \ref{sec:reg_chal}.

The ongoing challenge for privacy-preserving tools in permissionless solutions underscores the importance of developing decentralized technologies that protect human rights and liberties. While such tools cannot replace legal frameworks and democratic institutions, they can supplement them by providing privacy guarantees less vulnerable to business lobbying, governmental incompetence, and bad actors. Despite current limitations, blockchain systems show promise in developing tools based on trust-minimization, permissionless access, and cryptographic privacy.

\subsection{Incentives machine}
\label{subsec:manip}

The peculiar hyper-financialized nature of blockchain systems is very difficult to separate from technology, and there are good reasons why the idea of monetary rewards is so significant in blockchains. If we recall, blockchains are not just complex systems consisting of many functional layers, but also decentralised Peer-to-Peer (P2P) systems\footnote{Often communication layer in blockchain applications is labeled as P2P layer to differentiate it from the consensus layer. However, strictly speaking any permissionless decentralized system is a P2P system at high level.}. And until the appearance of Bitcoin one of the most difficult problems in decentralised systems was the problem of incentives for participants. 

Or put differently, why would participants of such systems want to contribute their time or resources to maintain it? Earlier generations of the decentralised systems known as P2P, the most well-known of which are torrents suffered from the same problem of free-riding. That is many people are happy to consume resources (e.g. download torrents), but few people want to contribute (e.g. seed torrents). If there are more people who consume than people who contribute, P2P system that may fail to maintain necessary properties such as liveness~\cite{feldman2005overcoming}. 

While P2P systems do work, their reliance on altruism means they struggle to (socially) scale effectively. The incentives in these systems are not scalable, as only those who value specific rewards, like social recognition, will contribute. What it means is that if we have a system which is based on a particular type of incentive e.g. social recognition, then it will attract contributions only from those who care about these specific rewards. But what if we want to attract resources and contributions from a wider range of participants? Bitcoin addressed this issue by introducing more universal monetary rewards (mining fees) for participants who contribute computational resources to maintain the system.

Ethereum expanded this concept by introducing smart contracts, enabling not just monetary transactions but also a wide range of decentralized applications (Dapps), extending the concept of monetary incentives further. These Dapps, including decentralized exchanges (DEXs) and various token-based digital assets, have given rise to a new field called Tokenomics. What started as a joke about token-based economics is now a legitimate research area~\cite{kensuke2024}. These developments demonstrate how decentralized systems can achieve "social scalability," a problem that earlier P2P systems, reliant on participant altruism, never fully resolved. New incentive mechanisms like gamification and social feedback loops, borrowed from social media, are also being integrated into these decentralized applications.

From a purely descriptive standpoint, these decentralized incentive mechanisms represent successful experiments in fostering participation and scalability. However, they also raise significant ethical concerns. The line between incentivization and manipulation becomes blurred when incentives influence participants' decisions and actions in ways that bypass their rational decision-making processes or exploit their psychological tendencies. The ethical issue arises when these incentives become manipulative, leading individuals to act in ways they might not have otherwise chosen. While these issues got some coverage in Human-Computer Interactions research under the label of "dark pattern"~\cite{gray_dark_2018}, they remain largely unexplored in blockchain space.

The transparency and autonomy often associated with blockchain technology do not necessarily mitigate these ethical concerns. Even in decentralized systems, the designers of protocols and Dapps hold considerable influence over user behavior. This power dynamic poses a moral challenge: while the goal may be to maintain and secure the network, the use of such incentive structures can cross the line from voluntary participation to manipulative coercion. Thus, the morality of these mechanisms depends not only on their outcomes but also on the ethical principles guiding their design and implementation.

\section{Obstacles for blockchain ethics and its limits}
\label{sec:obstacles}

As we have seen, blockchain ethics do deal with some multifaceted moral questions that concern not only individuals with blockchain applications but arguably the whole societies. So it stands to a reason that blockchain ethics should not only identify and explicate but also try to address these issues. This, however, is not something that should be taken for granted. As we know from other fields of technology ethics the impact of various ethical guidelines, ethical codes of conduct and other "paper ethics" is unfortunately rather limited~\cite{green_contestation_2021}~\cite{bietti_ethics_2020}. These initiatives are often perceived as extraneous, add-one unbinding frameworks imposed by institutions outside of the technical community~\cite{hagendorff_ethics_2020}. 

As a consequence, these initiatives either do not affect the decision-making of software developers or get integrated in a superficial manner as marketing or PR. Thus, we need to be aware of these specific challenges for blockchain ethics, and choose the practical strategies accordingly. 

We have briefly touched upon the topic of empirical grounding for ethics of technology in Section \ref{sec:eth_tech}. This issue is not only about the relation of high-level theories to empirical facts but also about the appropriate choice of conceptual frameworks. The first challenge here is a proper choice of the level of abstraction. Secondly, we have to avoid the pitfalls of conceptual confusion when trying to translate technological facts into non-technical arguments. Finally, we have to ensure that all insights that we obtain from our moral argumentation are relevant to the decision-makers, protocol designers, software engineers, and others who have the capacity to make a difference.

\subsection{Conceptual frameworks}

Blockchain systems indeed present non-trivial methodological challenges for the ethicists of technology. This issue can be referred to as a general problem of conceptual frameworks in interdisciplinary contexts~\cite{beers_eliciting_2009}. Different fields and areas of knowledge often have specific terminology and concepts. The same is true for blockchain research and engineering, which are based on concepts from cryptography, distributed systems, software engineering, and sometimes game theory. These specific terminologies can cause a lot of confusion due to polysemy and incorrect levels of abstraction.

A proficient language speaker can easily deal with the phenomenon of polysemy, where the same word can have completely different meanings, such as "bank" as a financial institution and "bank" as the edge of a river. Polysemy in scientific contexts, however, might require expert knowledge, for example, to distinguish between "field" in mathematics and "field" in physics. Blockchain terminology can often be misleading as well. One of the most infamous examples is the concept of "trust." Originally taken from cryptography in the narrow sense of Trusted-Third Party, it quickly became a central point for many confused interpretations, leading to (largely) misplaced arguments that blockchains can remove the need for trust between users completely or become a "trust machine"~\cite{economist2015promise}. The concept of "smart contract" is another good illustration of this problem~\cite{reyes2022emerging}.

Another related issue here is the problem of selecting the correct level of abstraction when choosing a model or providing a description of a complex system~\cite{floridi_method_2008}. Abstraction helps in simplifying complex systems by hiding unnecessary details and focusing on the essential aspects relevant to a particular context. This makes it easier to manage, understand, and work with complex systems without being overwhelmed by their intricacies. However, good abstractions are also critical for shared models or representations of systems that can be understood by different teams or individuals, ensuring everyone is on the same page regarding the system's design and functionality. For example, even within the blockchain industry, the understanding of "social consensus" is still in the process of being disentangled from ideas of consensus mechanisms~\cite{buterin2023blogpost}.

Given these challenges, it becomes imperative that blockchain ethics develops an interdisciplinary vocabulary that can avoid conceptual slippage-where terms are misunderstood or misapplied across disciplines-and selects appropriate levels of abstraction that align with the nuances of blockchain technology. Building such a vocabulary is not just a matter of semantics, it is essential for advancing ethical discourse in a way that is both informed and precise.

\subsection{Information asymmetries}
\label{subsec:inf_assym}

Finally, blockchain applications and blockchain ecosystems, in general, are characterized by information asymmetries. "Organic" information asymmetries stem from the fact that blockchain applications involve specialist knowledge from various expert fields: cryptography, distributed systems, economics. Thus, any analysis of blockchain applications always risks slipping into the area of highly specialized knowledge. However, the blockchain industry is also characterized by "intentional" information asymmetries rooted in monetary incentives~\cite{agarwal2023short}. Monetary rewards in current blockchain ecosystems are heavily skewed in favor of participants capable of leveraging their information advantage, which in turn creates many incentives at all levels to create and maintain these asymmetries.

Addressing "organic" asymmetries is arguably a more straightforward task that can benefit from the experience of interdisciplinary research and interdisciplinary epistemologies. Dealing with intentional information asymmetries may be more challenging but not fundamentally impossible. Some helpful insights could be found in adversarial risk analysis frameworks in cybersecurity, emphasizing the importance of validating information across multiple independent sources, cross-referencing data, and assessing the credibility of information sources. A good example of practical adversarial analysis of unethical behavior in the blockchain industry is presented by investigations of pseudonymous researcher ZackXBT\footnote{\url{https://x.com/zachxbt}}.

\subsection{Regulatory challenges}
\label{sec:reg_chal}

It is sometimes implicitly assumed that making blockchain solutions "more ethical" is the same thing as bringing them in line with regulatory compliance. This, however, is a misconception that stems from the 'ethics and compliance' label - an umbrella term covering corporate programs. In this context "ethics" is a label in the context of business management aimed at the development of internal corporate rules for proactive mitigation of regulatory risks.

It is important to keep in mind though that in general legal frameworks and ethical considerations, while often overlapping, operate on different principles. For a starter, there is no agreement within the legal theory itself that law should always be derived from and aligned with moral principles (the critical stance on alignment between morality and law comes from legal positivism). And while in practice most of the time law aims to reflect some generally accepted norms within the society, including moral ones, this is not necessarily always so. For instance, in authoritarian regimes law may simply reflect the self-interests of governing institutions.

Furthermore, even in an ideal scenario where the regulation of technologies is rooted in moral considerations, there is no complete overlap between law and ethics. For one, in the context of emerging technologies, there is a problem of legal lag, where efficiency and relevance of regulation are hampered by the disconnect between law and technological reality. So, for example, privacy protection regulation designed for centralized systems that prioritize control, accountability, and enforcement, can be inefficient or even counter-productive in the context of decentralized, pseudonymous systems~\cite{povse_data_2023}.

Another reason for the disconnect between the ethics of technology and regulation of technologies stems from the fact that even within the same jurisdiction there may be conflicting laws. For example, the General Data Protection Regulation (GDPR) in the EU emphasizes strict data privacy protections, while Know Your Customer (KYC) regulations require extensive personal data collection for anti-money laundering purposes. Furthermore, some KYC requirements might clash with fundamental human rights, such as the right to privacy under the European Convention on Human Rights, leading to legal ambiguities and compliance challenges~\cite{mitsilegas_evolving_2016}.

Thus, the ethical implications of regulating blockchain technologies are not always one-dimensional and may have widely different outcomes. Given that open source protocols and permissionless systems are transnational, ethical implications will vary greatly depending on the jurisdiction. So, for instance, in countries with authoritarian regimes, with laws that undermine human rights or are even patently inhumane, technologies may protect individuals offering privacy-preserving tools and censorship-resistant access to information, and protection of property in the form of permissionless digital currencies. In these context, the ability of such systems to resist regulatory enforcement can be considered morally desirable. 

However, even in democratic jurisdictions where the rule of law and human rights are respected, the same values that are built in decentralized, permissionless solutions can paradoxically conflict with regulations designed to ensure accountability, transparency, and oversight. Blockchain’s decentralized, pseudonymous nature challenges traditional regulatory mechanisms that rely on identifying individuals or controlling intermediaries, revealing a tension between the values embedded in blockchain and the priorities of legal frameworks even in societies committed to protecting human rights.

Market mechanisms may also present unexpected challenges at the crossroads of ethics and regulation of blockchain applications. As projects behind decentralized applications become financially successful, they morph into more traditional corporate entities that seek legal compliance and regulatory certainty. On one hand, these new businesses put pressure on regulators through lobbying efforts. On the other hand, these businesses also aim to influence the development of open-source blockchain protocols in line with their market interests. These processes may affect blockchain development on the ecosystem scale, and bring well-known effects of erosion of moral motivations (crowding out) under the pressure of economic interests~\cite{frey_cost_1997}. 

The dilemma posed by the conflict between blockchain’s embedded ethical values and regulatory demands raises a difficult question: Should ethical principles be compromised for compliance, or should regulations adapt to better align with these values? As blockchain technologies continue to evolve, some degree of regulatory harmonization is likely, particularly in areas where new technological realities demand a shift in legal approaches. For example, recognizing digital property rights and adapting legal frameworks for smart contracts may align regulations more closely with the decentralized nature of blockchain. However, ongoing conflicts, especially in areas like privacy, are likely to persist. Regulations that mandate transparency or user identification may remain at odds with blockchain’s privacy-preserving features, leading to persistent legal and ethical tensions that will need to be addressed.

\section{Practical ethics or science of Complex systems and their trade-offs}
\label{sec:pract_eth}

As we have seen from the previous sections, blockchain ethics can be considered a specific field of applied ethics with a unique scope and specific challenges. In Section \ref{sec:block_eth}, we also defined what we expect from blockchain ethics as a research tool. At a minimum, it should help us identify specific moral issues pertaining to the design and use of blockchain solutions, including instances of value conflicts, moral dilemmas, risks, harms, or other concerns. However, blockchain ethics is not necessarily limited to morally problematic issues; it should also help us understand how technology can promote human well-being and flourishing. Furthermore, as we discussed in Section \ref{sec:obstacles}, it should strive to be not just a research tool but also a feasible practice that can address the existing hurdles. Overcoming the challenges of conceptual confusion and information asymmetries is only the first step forward. We also need to look into specific tools that can bridge the gap between engineering practices and theoretical considerations.

\subsection{Ethical risk analysis}

Of particular interest here is the approach of ethical risk analysis~\cite{hansson_ethical_2017}. Originating from the practices of risk management, this approach aims to systematically address the ethical issues of risks. Of particular interest is the focus on the moral right of individuals to expose themselves to risk and whether others have the right to facilitate or contribute to this risk exposure. Analyzing roles and relationships between different stakeholders, it addresses the ethical boundaries between personal autonomy and external influence. This approach is fitting in the context of blockchain ethics analysis for several reasons.

Firstly, as discussed in Section \ref{subsec:manip}, complex incentive mechanisms blur the boundaries between autonomous decision-making and manipulative practices. Ethical risk analysis, by identifying the roles of beneficiaries, risk-exposed individuals, and decision-makers, can help ensure that incentive structures are transparent and that those who benefit from the risks are held accountable. Such analysis is useful for mapping out the complex web of decision-makers and understanding how their decisions impact others, for example, in the complex dynamics within Eigenlayer’s restaking mechanism~\cite{durvasula2024robustrestakingnetworks}, or conflicting incentives and risk distribution in different MEV (Miner Extractable Value) mitigation solutions~\cite{yang2024mev}.

Secondly, ethical risk analysis focuses on ways in which risk-relevant information is controlled, disseminated, or withheld. As discussed in Section \ref{subsec:inf_assym}, information asymmetry is one of the more persistent problems. Thus, it is crucial to identify and address these asymmetries by ensuring that all relevant risk-related information is disclosed and accessible to all participants. Systematic analysis of risk-related information dissemination also helps highlight cases of ethically problematic information flows not only between direct stakeholders but also within extended networks of independent information providers.\footnote{Good example of a project aiming to address information assymetries is L2Beat: \url{https://l2beat.com/scaling/summary}}.

Finally, ethical risk analysis is a practical approach that seeks to mitigate identified issues through the re-adjustment of the risk-benefit balance between stakeholders. For instance, we may seek to impose extra costs on risk-benefiting decision-makers or prevent persons or organizations from activities that would put them in the position of being risk-benefiting decision-makers.

\subsection{Value Sensitive Design}

Another practical approach of interest in the context of blockchain ethics is the method of Value-Sensitive Design (VSD), or Design for Values. It aims to integrate human values into the design process of technological systems from the outset, considering the ethical implications of technology alongside its functional requirements. VSD also provides methods for identifying relevant values, translating values into actionable design criteria (operationalization), and resolving emergent value conflicts~\cite{van_den_hoven_design_2015}.

This method is relevant for blockchain design and engineering for several reasons. For one, complex decentralized systems inevitably involve certain trade-offs in their designs, such as choices between security and performance. The same is true of blockchain systems, where the choice of architectures and implementations involves trade-offs~\cite{b_nasrulin_gromit_2022}~\cite{gramoli_diablo_2022}. But what is arguably particular for blockchain solutions is that, in addition to technological trade-offs, they often require us to consider  trade-offs between different types of non-technical goals and values.

As we have mentioned, some DAO and DeFi protocols get tempted to sacrifice values like decentralization or permissionless access for the sake of wider adoption. It is hard to always appreciate the long-term consequences of these choices, but these consequences are always present. For instance, these sacrifices can make systems less censorship-resistant or less secure, as centralized parties in these protocols become single points of failure. If many people rely on your system, this can have a serious ethical impact.

Another argument for the usefulness of value analysis in blockchain ethics is that often decentralized projects are driven by specific values. There are of course straightforward  cases like Monero cryptocurrency, which is largely a mono-value project designed and built for the value of privacy. It is also somewhat of an exception, since, unlike other values, privacy is well understood in different contexts, and thanks to robust research in cryptography and cybersecurity, we have a relatively good understanding of how privacy can be operationalized in technical terms. This is not the case, however, with many other blockchain solutions aim to achieve a wider set of values like Ethereum \cite{buterin2023blogpost}. Arguably, it is still not well understood how values like decentralization, censorship-resistance, auditability, and fairness could be operationalized and implemented in blockchain solutions. It stands to reason, then, that we could benefit from systematic methods that can help bring these values into reality.

\section{Conclusion}

This chapter has explored and addressed three key questions related to blockchain ethics. First, we examined the role of blockchain ethics within the broader context of technology ethics, identifying the goals and purposes that should guide ethical considerations in this field. Next, we outlined specific aspects of blockchain applications and decentralized technologies that define the uniqueness and scope of blockchain ethics, in the areas of permissionless solutions, incentive mechanisms, and privacy. We identified the key obstacles that blockchain ethics faces, including the development of correct conceptual models and the mitigation of information asymmetries. Finally, we argued that the most effective way to advance blockchain ethics is by approaching it as an engineering discipline, focusing on understanding and designing trade-offs in complex systems. Some applied approaches from the toolkit of applied ethics present particular interest here, namely ethical risk assessment (ERA) and Value Sensitive Design (VSD).

Blockchain technologies create a vibrant and somewhat unusual ecosystem of highly experimental platforms, digital services and products. But more importantly, the success of blockchain solutions has made a compelling point that decentralised computer systems can be a viable alternative to the centralised architectures that came to dominate our digitalised society. Permissionless decentralized systems are significant both from the technological and social perspectives, as they help to overcome fundamental bottlenecks of systems based on persistent identities and access control. And while it would be naive to think that software applications will be able to completely replace such critical institutions in our society as law or democratic institutions, they can certainly augment them. Ultimately, these architectures provide a design paradigm for  'anti-dystopian’ technologies where ethical norms, human rights and moral values are embedded in our crucial digital infrastructures.

\bibliography{references}


\begin{thebibliography}{43}
\ifx \bisbn   \undefined \def \bisbn  #1{ISBN #1}\fi
\ifx \binits  \undefined \def \binits#1{#1}\fi
\ifx \bauthor  \undefined \def \bauthor#1{#1}\fi
\ifx \batitle  \undefined \def \batitle#1{#1}\fi
\ifx \bjtitle  \undefined \def \bjtitle#1{#1}\fi
\ifx \bvolume  \undefined \def \bvolume#1{\textbf{#1}}\fi
\ifx \byear  \undefined \def \byear#1{#1}\fi
\ifx \bissue  \undefined \def \bissue#1{#1}\fi
\ifx \bfpage  \undefined \def \bfpage#1{#1}\fi
\ifx \blpage  \undefined \def \blpage #1{#1}\fi
\ifx \burl  \undefined \def \burl#1{\textsf{#1}}\fi
\ifx \doiurl  \undefined \def \doiurl#1{\url{https://doi.org/#1}}\fi
\ifx \betal  \undefined \def \betal{\textit{et al.}}\fi
\ifx \binstitute  \undefined \def \binstitute#1{#1}\fi
\ifx \binstitutionaled  \undefined \def \binstitutionaled#1{#1}\fi
\ifx \bctitle  \undefined \def \bctitle#1{#1}\fi
\ifx \beditor  \undefined \def \beditor#1{#1}\fi
\ifx \bpublisher  \undefined \def \bpublisher#1{#1}\fi
\ifx \bbtitle  \undefined \def \bbtitle#1{#1}\fi
\ifx \bedition  \undefined \def \bedition#1{#1}\fi
\ifx \bseriesno  \undefined \def \bseriesno#1{#1}\fi
\ifx \blocation  \undefined \def \blocation#1{#1}\fi
\ifx \bsertitle  \undefined \def \bsertitle#1{#1}\fi
\ifx \bsnm \undefined \def \bsnm#1{#1}\fi
\ifx \bsuffix \undefined \def \bsuffix#1{#1}\fi
\ifx \bparticle \undefined \def \bparticle#1{#1}\fi
\ifx \barticle \undefined \def \barticle#1{#1}\fi
\bibcommenthead
\ifx \bconfdate \undefined \def \bconfdate #1{#1}\fi
\ifx \botherref \undefined \def \botherref #1{#1}\fi
\ifx \url \undefined \def \url#1{\textsf{#1}}\fi
\ifx \bchapter \undefined \def \bchapter#1{#1}\fi
\ifx \bbook \undefined \def \bbook#1{#1}\fi
\ifx \bcomment \undefined \def \bcomment#1{#1}\fi
\ifx \oauthor \undefined \def \oauthor#1{#1}\fi
\ifx \citeauthoryear \undefined \def \citeauthoryear#1{#1}\fi
\ifx \endbibitem  \undefined \def \endbibitem {}\fi
\ifx \bconflocation  \undefined \def \bconflocation#1{#1}\fi
\ifx \arxivurl  \undefined \def \arxivurl#1{\textsf{#1}}\fi
\csname PreBibitemsHook\endcsname

\bibitem[\protect\citeauthoryear{Agarwal et~al.}{2023}]{agarwal2023short}
\begin{bchapter}
\bauthor{\bsnm{Agarwal}, \binits{S.}},
\bauthor{\bsnm{Atondo-Siu}, \binits{G.}},
\bauthor{\bsnm{Ordekian}, \binits{M.}},
\bauthor{\bsnm{Hutchings}, \binits{A.}},
\bauthor{\bsnm{Mariconti}, \binits{E.}},
\bauthor{\bsnm{Vasek}, \binits{M.}}:
\bctitle{Short paper: Defi deception—uncovering the prevalence of rugpulls in cryptocurrency projects}.
In: \bbtitle{International Conference on Financial Cryptography and Data Security},
pp. \bfpage{363}--\blpage{372}
(\byear{2023}).
\bcomment{Springer}
\end{bchapter}
\endbibitem

\bibitem[\protect\citeauthoryear{Abraham and Guegan}{2019}]{abraham2019other}
\begin{botherref}
\oauthor{\bsnm{Abraham}, \binits{L.}},
\oauthor{\bsnm{Guegan}, \binits{D.}}:
The other side of the coin: Risks of the libra blockchain.
arXiv preprint arXiv:1910.07775
(2019)
\end{botherref}
\endbibitem

\bibitem[\protect\citeauthoryear{Agerskov et~al.}{2023}]{agerskov_ethical_2023}
\begin{bchapter}
\bauthor{\bsnm{Agerskov}, \binits{S.}},
\bauthor{\bsnm{Pedersen}, \binits{A.B.}},
\bauthor{\bsnm{Beck}, \binits{R.}}:
\bctitle{Ethical guidelines for blockchain systems.}
(\byear{2023})
\end{bchapter}
\endbibitem

\bibitem[\protect\citeauthoryear{Beers and Bots}{2009}]{beers_eliciting_2009}
\begin{botherref}
\oauthor{\bsnm{Beers}, \binits{P.J.}},
\oauthor{\bsnm{Bots}, \binits{P.W.}}:
Eliciting conceptual models to support interdisciplinary research
\textbf{35}(3),
259--278
(2009)
\end{botherref}
\endbibitem

\bibitem[\protect\citeauthoryear{Blaze et~al.}{1999}]{blaze_role_1999}
\begin{botherref}
\oauthor{\bsnm{Blaze}, \binits{M.}},
\oauthor{\bsnm{Feigenbaum}, \binits{J.}},
\oauthor{\bsnm{Ioannidis}, \binits{J.}},
\oauthor{\bsnm{Keromytis}, \binits{A.D.}}:
The role of trust management in distributed systems security,
185--210
(1999)
\end{botherref}
\endbibitem

\bibitem[\protect\citeauthoryear{Boyd and Hargittai}{2010}]{boyd_facebook_2010}
\begin{botherref}
\oauthor{\bsnm{Boyd}, \binits{D.}},
\oauthor{\bsnm{Hargittai}, \binits{E.}}:
Facebook privacy settings: Who cares?
(2010)
\doiurl{10.5210/fm.v15i8.3086} .
Accessed 2025-02-28
\end{botherref}
\endbibitem

\bibitem[\protect\citeauthoryear{Bietti}{2020}]{bietti_ethics_2020}
\begin{bchapter}
\bauthor{\bsnm{Bietti}, \binits{E.}}:
\bctitle{From ethics washing to ethics bashing: a view on tech ethics from within moral philosophy}.
In: \bbtitle{Proceedings of the 2020 Conference on Fairness, Accountability, and Transparency},
pp. \bfpage{210}--\blpage{219}.
\bpublisher{{ACM}}, \blocation{???}
(\byear{2020}).
\doiurl{10.1145/3351095.3372860} .
\bcomment{Place: Barcelona Spain}.
\burl{https://dl.acm.org/doi/10.1145/3351095.3372860}
Accessed 2024-08-15
\end{bchapter}
\endbibitem

\bibitem[\protect\citeauthoryear{{B. Nasrulin} et~al.}{2022-08-15}]{b_nasrulin_gromit_2022}
\begin{bchapter}
\bauthor{\bsnm{{B. Nasrulin}}},
\bauthor{\bsnm{{M. De Vos}}},
\bauthor{\bsnm{{G. Ishmaev}}},
\bauthor{\bsnm{{J. Pouwelse}}}:
\bctitle{Gromit: Benchmarking the performance and scalability of blockchain systems}.
In: \bbtitle{2022 {IEEE} International Conference on Decentralized Applications and Infrastructures ({DAPPS})},
pp. \bfpage{56}--\blpage{63}
(\byear{2022-08-15}).
\doiurl{10.1109/DAPPS55202.2022.00015}
\end{bchapter}
\endbibitem

\bibitem[\protect\citeauthoryear{Buterin}{2023}]{buterin2023blogpost}
\begin{botherref}
\oauthor{\bsnm{Buterin}, \binits{V.}}:
Make Ethereum Cypherpunk Again.
Accessed: October 1, 2024
(2023).
\url{https://vitalik.eth.limo/general/2023/12/28/cypherpunk.html}
\end{botherref}
\endbibitem

\bibitem[\protect\citeauthoryear{Conlon et~al.}{2023}]{conlon_collapse_2023}
\begin{botherref}
\oauthor{\bsnm{Conlon}, \binits{T.}},
\oauthor{\bsnm{Corbet}, \binits{S.}},
\oauthor{\bsnm{Hu}, \binits{Y.}}:
The collapse of the {FTX} exchange: The end of cryptocurrency's age of innocence,
101277
(2023)
\doiurl{10.1016/j.bar.2023.101277} .
Accessed 2024-08-14
\end{botherref}
\endbibitem

\bibitem[\protect\citeauthoryear{Chaum}{1985}]{chaum_security_1985}
\begin{botherref}
\oauthor{\bsnm{Chaum}, \binits{D.}}:
Security without identification: transaction systems to make big brother obsolete
\textbf{28}(10),
1030--1044
(1985)
\doiurl{10.1145/4372.4373} .
Accessed 2024-08-15
\end{botherref}
\endbibitem

\bibitem[\protect\citeauthoryear{Chrisman}{2023}]{chrisman_what_2023}
\begin{bbook}
\bauthor{\bsnm{Chrisman}, \binits{M.}}:
\bbtitle{What Is this Thing Called Metaethics?}
\bpublisher{Routledge}, \blocation{???}
(\byear{2023})
\end{bbook}
\endbibitem

\bibitem[\protect\citeauthoryear{Cachin and Vukolić}{2017}]{cachin_blockchain_2017}
\begin{botherref}
\oauthor{\bsnm{Cachin}, \binits{C.}},
\oauthor{\bsnm{Vukolić}, \binits{M.}}:
Blockchain Consensus Protocols in the Wild.
{arXiv}
(2017).
\doiurl{10.48550/ARXIV.1707.01873} .
\url{https://arxiv.org/abs/1707.01873}
Accessed 2024-08-18
\end{botherref}
\endbibitem

\bibitem[\protect\citeauthoryear{Durvasula and Roughgarden}{2024}]{durvasula2024robustrestakingnetworks}
\begin{botherref}
\oauthor{\bsnm{Durvasula}, \binits{N.}},
\oauthor{\bsnm{Roughgarden}, \binits{T.}}:
Robust Restaking Networks
(2024).
\url{https://arxiv.org/abs/2407.21785}
\end{botherref}
\endbibitem

\bibitem[\protect\citeauthoryear{Deuber et~al.}{2022}]{deuber2022sok}
\begin{botherref}
\oauthor{\bsnm{Deuber}, \binits{D.}},
\oauthor{\bsnm{Ronge}, \binits{V.}},
\oauthor{\bsnm{R{\"u}ckert}, \binits{C.}}:
Sok: Assumptions underlying cryptocurrency deanonymizations.
Proceedings on Privacy Enhancing Technologies
(2022)
\end{botherref}
\endbibitem

\bibitem[\protect\citeauthoryear{Economist}{2015}]{economist2015promise}
\begin{barticle}
\bauthor{\bsnm{Economist}, \binits{T.}}:
\batitle{The promise of the blockchain: the trust machine}.
\bjtitle{The economist}
\bvolume{31},
\bfpage{27}
(\byear{2015})
\end{barticle}
\endbibitem

\bibitem[\protect\citeauthoryear{Feldman and Chuang}{2005}]{feldman2005overcoming}
\begin{barticle}
\bauthor{\bsnm{Feldman}, \binits{M.}},
\bauthor{\bsnm{Chuang}, \binits{J.}}:
\batitle{Overcoming free-riding behavior in peer-to-peer systems}.
\bjtitle{ACM sigecom exchanges}
\bvolume{5}(\bissue{4}),
\bfpage{41}--\blpage{50}
(\byear{2005})
\end{barticle}
\endbibitem

\bibitem[\protect\citeauthoryear{Floridi}{2008-09}]{floridi_method_2008}
\begin{botherref}
\oauthor{\bsnm{Floridi}, \binits{L.}}:
The method of levels of abstraction
\textbf{18}(3),
303--329
(2008-09)
\doiurl{10.1007/s11023-008-9113-7} .
Accessed 2024-08-18
\end{botherref}
\endbibitem

\bibitem[\protect\citeauthoryear{Frey and Oberholzer-Gee}{1997}]{frey_cost_1997}
\begin{botherref}
\oauthor{\bsnm{Frey}, \binits{B.S.}},
\oauthor{\bsnm{Oberholzer-Gee}, \binits{F.}}:
The cost of price incentives: An empirical analysis of motivation crowding-out
\textbf{87}(4),
746--755
(1997).
Publisher: {JSTOR}
\end{botherref}
\endbibitem

\bibitem[\protect\citeauthoryear{Gramoli et~al.}{2022}]{gramoli_diablo_2022}
\begin{bchapter}
\bauthor{\bsnm{Gramoli}, \binits{V.}},
\bauthor{\bsnm{Guerraoui}, \binits{R.}},
\bauthor{\bsnm{Lebedev}, \binits{A.}},
\bauthor{\bsnm{Natoli}, \binits{C.}},
\bauthor{\bsnm{Voron}, \binits{G.}}:
\bctitle{Diablo: A benchmark suite for blockchains}.
\bpublisher{Zenodo}, \blocation{???}
(\byear{2022}).
\doiurl{10.5281/ZENODO.7707312} .
\burl{https://zenodo.org/record/7707312}
Accessed 2024-08-18
\end{bchapter}
\endbibitem

\bibitem[\protect\citeauthoryear{Gray et~al.}{2018}]{gray_dark_2018}
\begin{bchapter}
\bauthor{\bsnm{Gray}, \binits{C.M.}},
\bauthor{\bsnm{Kou}, \binits{Y.}},
\bauthor{\bsnm{Battles}, \binits{B.}},
\bauthor{\bsnm{Hoggatt}, \binits{J.}},
\bauthor{\bsnm{Toombs}, \binits{A.L.}}:
\bctitle{The dark (patterns) side of {UX} design},
pp. \bfpage{1}--\blpage{14}
(\byear{2018})
\end{bchapter}
\endbibitem

\bibitem[\protect\citeauthoryear{Green}{2021}]{green_contestation_2021}
\begin{botherref}
\oauthor{\bsnm{Green}, \binits{B.}}:
The contestation of tech ethics: A sociotechnical approach to technology ethics in practice
\textbf{2}(3),
209--225
(2021)
\doiurl{10.23919/JSC.2021.0018} .
Accessed 2024-08-15
\end{botherref}
\endbibitem

\bibitem[\protect\citeauthoryear{Hagendorff}{2020}]{hagendorff_ethics_2020}
\begin{botherref}
\oauthor{\bsnm{Hagendorff}, \binits{T.}}:
The ethics of {AI} ethics: An evaluation of guidelines
\textbf{30}(1),
99--120
(2020)
\doiurl{10.1007/s11023-020-09517-8} .
Accessed 2024-08-15
\end{botherref}
\endbibitem

\bibitem[\protect\citeauthoryear{Hansson}{2017a}]{hansson_ethical_2017}
\begin{botherref}
\oauthor{\bsnm{Hansson}, \binits{S.O.}}:
Ethical risk analysis,
157--172
(2017)
\end{botherref}
\endbibitem

\bibitem[\protect\citeauthoryear{Hansson}{2017b}]{hansson_theories_2017}
\begin{botherref}
\oauthor{\bsnm{Hansson}, \binits{S.O.}}:
Theories and methods for the ethics of technology,
1--14
(2017)
\end{botherref}
\endbibitem

\bibitem[\protect\citeauthoryear{Hansson}{2023}]{hansson_moral_2023}
\begin{botherref}
\oauthor{\bsnm{Hansson}, \binits{S.O.}}:
Moral philosophy has much more to offer
\textbf{43}(2),
238--239
(2023)
\doiurl{10.1111/risa.13918} .
Accessed 2024-08-15
\end{botherref}
\endbibitem

\bibitem[\protect\citeauthoryear{Kensuke}{2024}]{kensuke2024}
\begin{bchapter}
\bauthor{\bsnm{Kensuke}, \binits{I.T.O.}}:
\bctitle{Cryptoeconomics and tokenomics as economics: A survey with opinions}.
In: \bbtitle{2024 IEEE International Conference on Blockchain and Cryptocurrency (ICBC)},
pp. \bfpage{729}--\blpage{746}
(\byear{2024}).
\doiurl{10.1109/ICBC59979.2024.10634383}
\end{bchapter}
\endbibitem

\bibitem[\protect\citeauthoryear{Laine et~al.}{2024}]{laine_ethics-based_2024}
\begin{botherref}
\oauthor{\bsnm{Laine}, \binits{J.}},
\oauthor{\bsnm{Minkkinen}, \binits{M.}},
\oauthor{\bsnm{Mäntymäki}, \binits{M.}}:
Ethics-based {AI} auditing: A systematic literature review on conceptualizations of ethical principles and knowledge contributions to stakeholders
\textbf{61}(5),
103969
(2024)
\doiurl{10.1016/j.im.2024.103969} .
Accessed 2024-08-14
\end{botherref}
\endbibitem

\bibitem[\protect\citeauthoryear{Munn}{2023}]{munn_uselessness_2023}
\begin{botherref}
\oauthor{\bsnm{Munn}, \binits{L.}}:
The uselessness of {AI} ethics
\textbf{3}(3),
869--877
(2023)
\doiurl{10.1007/s43681-022-00209-w} .
Accessed 2024-08-14
\end{botherref}
\endbibitem

\bibitem[\protect\citeauthoryear{Mitsilegas and Vavoula}{2016}]{mitsilegas_evolving_2016}
\begin{botherref}
\oauthor{\bsnm{Mitsilegas}, \binits{V.}},
\oauthor{\bsnm{Vavoula}, \binits{N.}}:
The evolving {EU} anti-money laundering regime: challenges for fundamental rights and the rule of law
\textbf{23}(2),
261--293
(2016).
Publisher: {SAGE} Publications Sage {UK}: London, England
\end{botherref}
\endbibitem

\bibitem[\protect\citeauthoryear{Povše et~al.}{2023}]{povse_data_2023}
\begin{bchapter}
\bauthor{\bsnm{Povše}, \binits{D.F.}},
\bauthor{\bsnm{Favenza}, \binits{A.}},
\bauthor{\bsnm{Frey}, \binits{D.}},
\bauthor{\bsnm{Mann}, \binits{Z.A.}},
\bauthor{\bsnm{Palomares}, \binits{A.}},
\bauthor{\bsnm{Piatti}, \binits{L.}},
\bauthor{\bsnm{Schroers}, \binits{J.}}:
\bctitle{Data protection challenges in distributed ledger and blockchain technologies: A combined legal and technical analysis}.
In: \bbtitle{Building Cybersecurity Applications with Blockchain and Smart Contracts},
pp. \bfpage{127}--\blpage{152}.
\bpublisher{Springer}, \blocation{???}
(\byear{2023})
\end{bchapter}
\endbibitem

\bibitem[\protect\citeauthoryear{Qi et~al.}{2024}]{qi2024sok}
\begin{barticle}
\bauthor{\bsnm{Qi}, \binits{H.}},
\bauthor{\bsnm{Xu}, \binits{M.}},
\bauthor{\bsnm{Yu}, \binits{D.}},
\bauthor{\bsnm{Cheng}, \binits{X.}}:
\batitle{Sok: Privacy-preserving smart contract}.
\bjtitle{High-Confidence Computing}
\bvolume{4}(\bissue{1}),
\bfpage{100183}
(\byear{2024})
\doiurl{10.1016/j.hcc.2023.100183}
\end{barticle}
\endbibitem

\bibitem[\protect\citeauthoryear{Reyes}{2022}]{reyes2022emerging}
\begin{botherref}
\oauthor{\bsnm{Reyes}, \binits{C.L.}}:
Emerging technology's language wars: Smart contracts.
Wisconsin Law Review Forward,
85
(2022)
\end{botherref}
\endbibitem

\bibitem[\protect\citeauthoryear{Swartz}{2018}]{swartz2018bitcoin}
\begin{barticle}
\bauthor{\bsnm{Swartz}, \binits{L.}}:
\batitle{What was bitcoin, what will it be? the techno-economic imaginaries of a new money technology}.
\bjtitle{Cultural studies}
\bvolume{32}(\bissue{4}),
\bfpage{623}--\blpage{650}
(\byear{2018})
\end{barticle}
\endbibitem

\bibitem[\protect\citeauthoryear{Troncoso et~al.}{2022}]{troncoso_deploying_2022}
\begin{botherref}
\oauthor{\bsnm{Troncoso}, \binits{C.}},
\oauthor{\bsnm{Bogdanov}, \binits{D.}},
\oauthor{\bsnm{Bugnion}, \binits{E.}},
\oauthor{\bsnm{Chatel}, \binits{S.}},
\oauthor{\bsnm{Cremers}, \binits{C.}},
\oauthor{\bsnm{Gürses}, \binits{S.}},
\oauthor{\bsnm{Hubaux}, \binits{J.-P.}},
\oauthor{\bsnm{Jackson}, \binits{D.}},
\oauthor{\bsnm{Larus}, \binits{J.R.}},
\oauthor{\bsnm{Lueks}, \binits{W.}},
\oauthor{\bsnm{Oliveira}, \binits{R.}},
\oauthor{\bsnm{Payer}, \binits{M.}},
\oauthor{\bsnm{Preneel}, \binits{B.}},
\oauthor{\bsnm{Pyrgelis}, \binits{A.}},
\oauthor{\bsnm{Salathé}, \binits{M.}},
\oauthor{\bsnm{Stadler}, \binits{T.}},
\oauthor{\bsnm{Veale}, \binits{M.}}:
Deploying decentralized, privacy-preserving proximity tracing
\textbf{65}(9),
48--57
(2022)
\doiurl{10.1145/3524107} .
Accessed 2024-08-15
\end{botherref}
\endbibitem

\bibitem[\protect\citeauthoryear{Timmons}{2012}]{timmons_moral_2012}
\begin{bbook}
\bauthor{\bsnm{Timmons}, \binits{M.}}:
\bbtitle{Moral Theory: An Introduction}.
\bpublisher{Rowman \& Littlefield Publishers}, \blocation{???}
(\byear{2012})
\end{bbook}
\endbibitem

\bibitem[\protect\citeauthoryear{Unger et~al.}{2015}]{unger2015sok}
\begin{bchapter}
\bauthor{\bsnm{Unger}, \binits{N.}},
\bauthor{\bsnm{Dechand}, \binits{S.}},
\bauthor{\bsnm{Bonneau}, \binits{J.}},
\bauthor{\bsnm{Fahl}, \binits{S.}},
\bauthor{\bsnm{Perl}, \binits{H.}},
\bauthor{\bsnm{Goldberg}, \binits{I.}},
\bauthor{\bsnm{Smith}, \binits{M.}}:
\bctitle{Sok: secure messaging}.
In: \bbtitle{2015 IEEE Symposium on Security and Privacy},
pp. \bfpage{232}--\blpage{249}
(\byear{2015}).
\bcomment{IEEE}
\end{bchapter}
\endbibitem

\bibitem[\protect\citeauthoryear{Van~den Hoven}{2008}]{van_den_hoven_moral_2008}
\begin{botherref}
\oauthor{\bsnm{Hoven}, \binits{J.}}:
Moral methodology and information technology,
49--67
(2008)
\end{botherref}
\endbibitem

\bibitem[\protect\citeauthoryear{Van~den Hoven et~al.}{2015}]{van_den_hoven_design_2015}
\begin{botherref}
\oauthor{\bsnm{Hoven}, \binits{J.}},
\oauthor{\bsnm{Vermaas}, \binits{P.E.}},
\oauthor{\bsnm{Poel}, \binits{I.}}:
Design for values: An introduction,
1--7
(2015)
\end{botherref}
\endbibitem

\bibitem[\protect\citeauthoryear{Van~Dijck et~al.}{2018}]{van_dijck_platform_2018}
\begin{bbook}
\bauthor{\bsnm{Van~Dijck}, \binits{J.}},
\bauthor{\bsnm{Poell}, \binits{T.}},
\bauthor{\bsnm{De~Waal}, \binits{M.}}:
\bbtitle{The Platform Society: Public Values in a Connective World}.
\bpublisher{Oxford university press}, \blocation{???}
(\byear{2018})
\end{bbook}
\endbibitem

\bibitem[\protect\citeauthoryear{Van~de Poel and Royakkers}{2023}]{van_de_poel_ethics_2023}
\begin{bbook}
\bauthor{\bsnm{Poel}, \binits{I.}},
\bauthor{\bsnm{Royakkers}, \binits{L.}}:
\bbtitle{Ethics, Technology, and Engineering: An Introduction}.
\bpublisher{John Wiley \& Sons}, \blocation{???}
(\byear{2023})
\end{bbook}
\endbibitem

\bibitem[\protect\citeauthoryear{Yang et~al.}{2024}]{yang2024mev}
\begin{bchapter}
\bauthor{\bsnm{Yang}, \binits{S.}},
\bauthor{\bsnm{Zhang}, \binits{F.}},
\bauthor{\bsnm{Huang}, \binits{K.}},
\bauthor{\bsnm{Chen}, \binits{X.}},
\bauthor{\bsnm{Yang}, \binits{Y.}},
\bauthor{\bsnm{Zhu}, \binits{F.}}:
\bctitle{Sok: Mev countermeasures}.
In: \bbtitle{Proceedings of the Workshop on Decentralized Finance and Security}.
\bsertitle{DeFi '24},
pp. \bfpage{21}--\blpage{30}.
\bpublisher{Association for Computing Machinery},
\blocation{New York, NY, USA}
(\byear{2024}).
\doiurl{10.1145/3689931.3694911} .
\burl{https://doi.org/10.1145/3689931.3694911}
\end{bchapter}
\endbibitem

\bibitem[\protect\citeauthoryear{Zhang et~al.}{2019}]{zhang2019sok}
\begin{botherref}
\oauthor{\bsnm{Zhang}, \binits{R.}},
\oauthor{\bsnm{Xue}, \binits{R.}},
\oauthor{\bsnm{Liu}, \binits{L.}}:
Security and privacy on blockchain.
ACM Comput. Surv.
\textbf{52}(3)
(2019)
\doiurl{10.1145/3316481}
\end{botherref}
\endbibitem

\end{thebibliography}

\end{document}